\documentclass[%
 reprint,
 superscriptaddress,
 amsmath,amssymb,
 aps,
prl,
]{revtex4-2}

\usepackage{graphicx}%
\usepackage{dcolumn}%

\usepackage{hyperref}
\usepackage[dvipsnames]{xcolor}
\hypersetup{
  colorlinks=true,
  hypertexnames=false,
  linktocpage,
  colorlinks=true, 
  urlcolor=magenta!90!black,    
  linkcolor=blue!60!black, 
  citecolor=black!60 
}%

\usepackage{amsthm,bm}
\usepackage{braket}
\usepackage{physics}
\usepackage{tikz}
\usetikzlibrary{quantikz}

\usepackage[ruled, vlined, linesnumbered]{algorithm2e}
\usepackage[percent]{overpic}

\usepackage{xprintlen}
\usepackage{layouts}
\usepackage{comment}
\usepackage[capitalize]{cleveref}
\theoremstyle{definition}

\begin{document}
\title{A robust phase of continuous transversal gates in quantum stabilizer codes}
\author{Eric Huang}
 \affiliation{
 Joint Center for Quantum Information and Computer Science, NIST/University of Maryland,
College Park, Maryland 20742, USA.
}
\author{Pierre-Gabriel Rozon}
\affiliation{Physics Department, McGill University, Montr\'eal, Qu\'ebec H3A 2T8, Canada.}
\author{Arpit Dua}
\affiliation{Department of Physics, Virginia Tech, Blacksburg, Virginia 24061, USA.}
\author{Sarang Gopalakrishnan}
\affiliation{Princeton Quantum Initiative, Princeton University, Princeton, New Jersey 08540, USA.}
\affiliation{Department of Electrical and Computer Engineering,
Princeton University, Princeton, New Jersey 08540, USA}
\author{Michael J. Gullans}
 \affiliation{
 Joint Center for Quantum Information and Computer Science, NIST/University of Maryland,
College Park, Maryland 20742, USA.
}
\affiliation{National Institute of Standards and Technology, Gaithersburg, MD 20899, USA.}
\begin{abstract}
    A quantum error correcting code protects encoded logical information against
    errors.
    Transversal gates are a naturally fault-tolerant way to manipulate logical
    qubits but cannot be universal themselves.
    Protocols such as magic state distillation are needed to achieve
    universality via measurements and postselection.
    A phase is a region of parameter space with smoothly varying large-scale
    statistical properties except at its boundaries.
    Here, we find a phase of continuously tunable logical unitaries for the
    surface code implemented by transversal operations and decoding that
    is robust against dephasing errors.
    The logical unitaries in this phase have an infidelity that is exponentially
    suppressed in the code distance compared to their rotation angles.
    We exploit this to design a simple fault-tolerant protocol for
    continuous-angle logical rotations.
    This lowers the overhead for applications requiring many
    small-angle rotations such as quantum simulation.
\end{abstract}
\date{\today}
\maketitle

\emph{Introduction}---%
Fault-tolerant quantum computing requires
a universal set of fault-tolerant logical gates.
Transversal unitary gates do not spread errors within code blocks so they are
naturally fault-tolerant,
but the Eastin-Knill theorem%
~\cite{eastinRestrictionsTransversalEncoded2009}
rules out a universal set of transversal gates.

The conventional strategy for fault-tolerant non-transversal gates
is to produce magic states for injection by gate teleportation%
~\cite{gottesmanDemonstratingViabilityUniversal1999}.
This requires mid-circuit measurement and feed-forward,
which has been demonstrated experimentally in linear optics%
~\cite{pittmanDemonstrationFeedforwardControl2002,prevedelHighspeedLinearOptics2007,waltherExperimentalOnewayQuantum2005},
trapped ions%
~\cite{riebeDeterministicQuantumTeleportation2004,barrettDeterministicQuantumTeleportation2004},
superconducting circuits%
~\cite{risteFeedbackControlSolidState2012,steffenDeterministicQuantumTeleportation2013},
and
neutral atoms%
~\cite{singhMidcircuitCorrectionCorrelated2023,huieRepetitiveReadoutRealTime2023,bluvsteinLogicalQuantumProcessor2024}.
Magic state distillation produces high fidelity magic states from many lower
quality ones%
~\cite{bravyiUniversalQuantumComputation2005}.
Numerous refinements have reduced its resource requirements%
~\cite{meierMagicstateDistillationFourqubit2012,bravyiMagicstateDistillationLow2012,websterReducingOverheadQuantum2015,gidneyEfficientMagicState2019,litinskiMagicStateDistillation2019,itogawaEvenMoreEfficient2024},
while alternative methods
of growing magic states with syndrome extraction and postselection 
~\cite{liMagicStateFidelity2015,singhHighfidelityMagicstatePreparation2022,bombinFaultTolerantPostselectionLowOverhead2024}
have culminated in magic state cultivation%
~\cite{gidneyMagicStateCultivation2024,vakninEfficientMagicState2025,chenEfficientMagicState2025,sahayFoldtransversalSurfaceCode2025,claesCultivatingStatesSurface2025}.

Although distillation schemes may be optimized for small rotation
angles~\cite{campbellEfficientMagicState2016},
most protocols produce fixed-angle rotations so synthesizing arbitrary angle
rotations up to some tolerance
$\epsilon$
still requires
$O(\log(1/\epsilon))$
operations~\cite{rossOptimalAncillafreeClifford+T2016}.
This becomes costly for small rotations where we want $\epsilon$ to be much
smaller than the rotation.
Thus non-distillation schemes for continuous angles based on rotations and
measurements with postselection have gained interest%
~\cite{choiFaultTolerantNonClifford2023,heHighfidelityInitializationLogical2025}.
Practical limitations of near-term devices have also motivated
partially fault-tolerant analog rotations%
~\cite{gavrielTransversalInjectionDirect2023,akahoshiPartiallyFaultTolerantQuantum2024,toshioPracticalQuantumAdvantage2024,ismailTransversalSTARArchitecture2025}.

In this Letter, we uncover a new phase of continuous-angle logical unitaries
in surface codes subjected to transversal coherent rotations%
~\cite{bravyiCorrectingCoherentErrors2018}.
We observe that these unitaries are robust against dephasing in that
the applied logical unitary is known up to a logical dephasing rate $q$
that is exponentially smaller than its logical rotation angle $\phi$ as the code
distance $d$ increases, so that on average $q/|\phi| \sim e^{-\kappa d}$
for some constant $\kappa>0$.
We apply this robustness to design a new fault-tolerant protocol that
implements logical rotations using only transversal coherent rotations, syndrome
measurements, and decoder-assisted corrections.
We then leverage repeated rounds of these logical rotations with adaptive
feed-forward to avoid postselection,
resulting in an efficient fault-tolerant protocol for preparing continuously
tunable small-angle magic states.

\begin{figure}[t]
    \begin{overpic}[width=\linewidth]{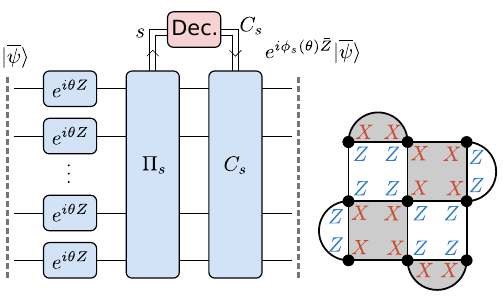}
        \put(0, 55){\textbf{(a)}}
        \put(63, 35){\textbf{(c)}}
    \end{overpic}
    \begin{overpic}[width=\linewidth]{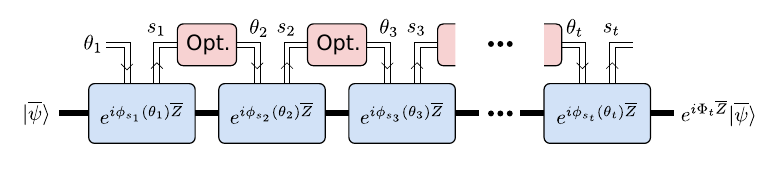}
        \put(0, 15){\textbf{(b)}}
    \end{overpic}
    \caption{(a) A CSS code state $\ket{\overline{\psi}}$ subjected to uniform
    coherent $Z$ rotations by $\theta$,
    followed by projective measurement of $X$-checks $\Pi_s$ to give a  syndrome
    $s$
    and application of the correction $C_s$ from the decoder
    results in a logical channel that is a known coherent $Z$
    rotation by angle $\phi_s(\theta)$.
    (b) Repeating $t$ rounds of coherent rotation by optimally chosen
    physical rotation angles $\theta_j$
    with error correction implements a potentially non-Clifford logical
    rotation by $\Phi_t=\theta_1+\cdots + \theta_t$.
    The classical angle optimizer (Opt.) is shaded in pink.
    (c) The $d=3$ square rotated surface code
    with qubits on vertices, $X$-checks on dark faces,
    and $Z$-checks generators on light faces.
    }
    \label{fig:setup}
\end{figure}

\emph{Background}---%
The square rotated surface code is an $[[ n, k, d ]]$
Calderbank-Shor-Steane (CSS) code which encodes $k=1$ logical qubit over $n=d^2$ physical
qubits for odd distance $d\ge 3$%
~\cite{wenQuantumOrdersExact2003,bombinOptimalResourcesTopological2007,fowlerSurfaceCodesPractical2012}.
This variant of the Kitaev surface code%
~\cite{kitaevFaulttolerantQuantumComputation2003,bravyiQuantumCodesLattice1998}
has been the most popular in surface code experiments%
~\cite{krinnerRealizingRepeatedQuantum2022,zhaoRealizationErrorCorrectingSurface2022,acharyaSuppressingQuantumErrors2023,acharyaQuantumErrorCorrection2024,berthusenExperimentsFourdimensionalSurface2024,bluvsteinArchitecturalMechanismsUniversal2025}.
The qubits live on the vertices of a $d\times d$ square grid
whose faces are shaded light and dark in a checkerboard pattern,
with $X$-type and $Z$-type stabilizer generators on dark and light faces
respectively, as depicted in Fig.~\ref{fig:setup}(c).
These local checks are weight-4 in the bulk and weight-2 on the boundary.
A code state can be fault-tolerantly prepared from a product state
by measuring stabilizer generators and correcting or,
alternatively, by using a linear-depth unitary circuit with flag
qubits~\cite{zenQuantumCircuitDiscovery2024}.

For storage with perfect measurements,
it is resilient against dephasing up to an optimal threshold rate of
$p\approx 11\%$~\cite{dennisTopologicalQuantumMemory2002}
corresponding to the phase boundary of a random-bond Ising model.
Coherent errors are systematic unitary rotations such as $\exp(i\theta Z)$ with
rotation angle $\theta\in(-\pi/2, \pi/2]$.
Although coherent errors acting on CSS codes are difficult to analyze in general%
~\cite{huangPerformanceQuantumError2019,iversonCoherenceLogicalQuantum2020,suzukiEfficientSimulationQuantum2017},
\textcite{bravyiCorrectingCoherentErrors2018} used a
Majorana free-fermion method to estimate a threshold $Z$-rotation angle of
$\theta_c\approx 0.11\pi$ for the surface code,
which has been confirmed by mappings to complex statistical
mechanics models~\cite{behrendsSurfaceCodesQuantum2022,vennCoherentErrorThreshold2022,behrendsStatisticalMechanicalMapping2024,behrendsSurfaceCodePauli2025}.
Applying coherent errors followed by syndrome measurement gives rise to phases of high
magic~\cite{niroulaPhaseTransitionMagic2024},
which motivates intentionally applying coherent rotations to produce magic
states.

\emph{Logical coherent rotation and dephasing}---%
Consider a round of transversal coherent rotations and decoding as shown in
Fig.~\ref{fig:setup}(a).
A code state is transversally rotated by
$U_\theta:={[\exp(i\theta Z)]}^{\otimes n}$
after which a syndrome $s$ is measured.
The correction $C_s$ from the decoder is applied,
returning the system to the code space.
If the CSS code has even-weight stabilizers with odd $X$ and $Z$ distances and
$U_\theta$ commutes with the time-reversal transformation,
then the resulting logical operation is  unitary~\cite{chengEmergentUnitaryDesigns2024}.
These conditions are satisfied in square rotated odd-distance surface codes,
resulting in a logical unitary
$\exp(i\phi_s \overline{Z})$
whose logical rotation angle $\phi_s(\theta)$ depends only on $\theta$ and
$s$.  More precisely, we have the identity \cite{bravyiCorrectingCoherentErrors2018}
\begin{equation} \label{eqn:proj}
\Pi_0 C_s U_\theta \Pi_{0} = \sqrt{p(s)} \, \Pi_0  e^{i\phi_s(\theta) \overline{Z}},
\end{equation}
where $\Pi_0$ is the projector onto the trivial $X$-syndrome, $C_s$ is a Pauli
$Z$-operator that produces $X$-syndrome $s$ and $p(s)$ is the syndrome
measurement probability.
Without noise, the applied logical angle is random but known with certainty.

When this logical rotation is repeated for $t$ rounds, as shown in
Fig.~\ref{fig:setup}(b), with optimally chosen rotation angles $\theta_j$ for
each round $j$,
then the logical rotation angles will sum to a total logical rotation
angle $\Phi_t$ which can be steered towards to a non-Clifford
target logical angle $\Phi_T$.

It thus remains to show the fault-tolerance of the protocol. $X$ errors
on the input state or before the first round of $X$ syndrome
measurements merely cause logical $X$ errors that do not affect the logical $Z$
angle $\phi_s$ due to the identity in Eq.~(\ref{eqn:proj}).
The protocol can also be made fault-tolerant against
measurement faults and Pauli errors that occur after the first round of syndrome
measurement.
To see this note that the syndrome extraction projects the state
into a syndrome subspace,
fixing the coherent rotation angle up to a Pauli error,
so to be robust against these errors we just need to determine which syndrome subspace the system was in at the first round.  This initial value of the syndrome can be
 learned by using $O(d)$ rounds of conventional syndrome extraction (see Appendix)
~\cite{dennisTopologicalQuantumMemory2002},
or by other means~\cite{zhouAlgorithmicFaultTolerance2024}.
On the other hand, since $Z$ errors on the input state, or following the coherent rotation~\cite{mckayEfficientGatesQuantum2017}, cause syndromes from the first round of syndrome extraction that  are indistinguishable from coherent $Z$
rotations, 
the applied logical rotation angle is no longer knowable with certainty in the presence of $Z$ noise.
The effect of dephasing hence warrants a more careful study, which is the focus of this work.

As a general model for incoherent $Z$ noise, we consider a round of the protocol where each qubit also suffers
dephasing at a known rate $p$ during the coherent rotations,
so that each physical qubit is subjected to the channel~\cite{huangPerformanceQuantumError2019}
\begin{align}\label{eqn:physicalchannel}
    \mathcal{E}(\rho) &=
    e^{i\theta Z}[(1 - p) \rho + p Z \rho Z] e^{-i\theta Z}.
\end{align}
After extracting a syndrome $s$ and decoding,
the logical channel will be of the same form
\begin{align}
    \overline{\mathcal{E}}_s(\rho) &=
    e^{i\phi_s Z}[(1 - q_s) \rho + q_s Z \rho Z] e^{-i\phi_s Z}%
    \label{eqn:logicalchannel}
\end{align}
but with a different logical dephasing rate $q_s$ and a logical rotation angle
$\phi_s$.
These are syndrome-dependent functions of the physical qubit channel parameters
$(p, \theta)$ and may be calculated using the tensor network
method of \textcite{darmawanTensorNetworkSimulationsSurface2017}.
The parameters will depend on the specific decoder,
but in this work
we use the PyMatching decoder~\cite{higgottSparseBlossomCorrecting2023}.
Although other decoders may be better-tailored to handle coherent errors%
~\cite{darmawanLineartimeGeneralDecoding2018},
the results will qualitatively be the same.

The $\theta=0$ case corresponds to dephasing only,
with the logical rotation angle $\phi_s=0$ for all syndromes $s$.
In the low-$p$ and small $\theta$ regime,
the most common syndrome is the trivial syndrome $\vec{0}$
so the $\phi_s$ distribution is peaked at the angle $\phi_0$ corresponding to the
trivial syndrome.
Above the error correction threshold,
$\phi_s$ becomes almost uniformly distributed.
More properties of this distribution are listed in the Appendix.

\begin{figure*}[t]
    \begin{center}
        \begin{overpic}[scale=1.0]{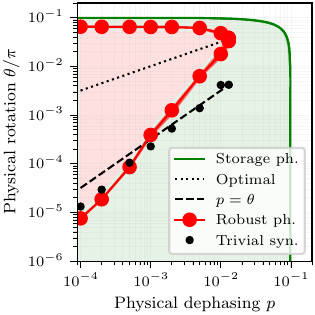}
            \put(0, 100){\textbf{(a)}}
        \end{overpic}
        \hspace{5mm}
        \begin{overpic}[scale=1.0]{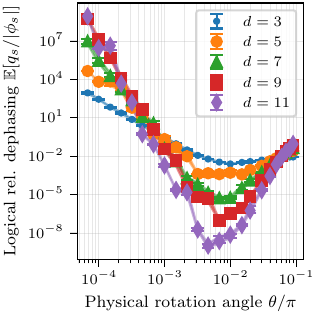}
            \put(0, 100){\textbf{(b)}}
        \end{overpic}
        \hspace{5mm}
        \begin{overpic}[scale=1.0]{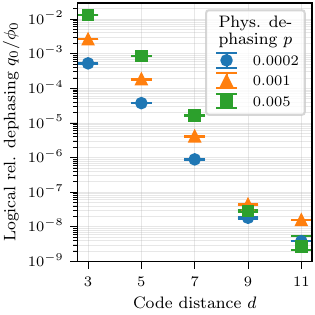}
            \put(0, 100){\textbf{(c)}}
        \end{overpic}
        \vspace{-5mm}
    \end{center}
    \caption{(a) Phase diagram in physical chanel parameter space $(p,\theta)$.
    In the robust logical coherent phase (shaded pink),
    the mean relative dephasing $\mathbb{E}[q_s/|\phi_s|]\to 0$
    as $d\to\infty$ using the PyMatching
    decoder~\cite{higgottSparseBlossomCorrecting2023}.
    It is contained within the correctable phase for conventional QEC storage
    (shaded green) where
    the diamond distance from the identity channel
    $\left\|\overline{\mathcal{E}}\ - \mathcal{I}\right\|_\diamond\to 0$
    as $d\to\infty$.
    (b) For a fixed physical dephasing $p=0.001$, the mean relative dephasing
    has a minimum value at a $\theta$ in the robust phase that is bounded from
    above and below by phase transitions.
    (c) The logical relative dephasing at the target angle that succeeds with 50\%
    probability is exponentially suppressed with the distance of the code.}
    \label{fig:phase}
\end{figure*}

\emph{Robust Phase}---%
In principle, a coherent rotation by a known angle $\phi_s$ is reversible,
unlike dephasing by $q_s$,
which destroys information and increases entropy.
But since each $\phi_s$ is of a random magnitude,
the logical \emph{relative dephasing} $q_s/|\phi_s|$
is the relevant measure of decoherence in the final state.
We find evidence for the existence a region of $(p,\theta)$ space called
the \emph{robust phase} where
the mean logical relative dephasing $\mathbb{E}[q_s/|\phi_s|]$ over the syndrome
distribution $p(s)$ is exponentially suppressed in the thermodynamic limit
$d\to\infty$.
This phase is highlighted in Fig.~\ref{fig:phase}(a)
but we would expect its precise boundary to shift with a more optimal
decoder.
For a fixed $p$,
$\theta$ values in this phase are bounded from above and below by two phase
transitions.
The upper critical $\theta$ beyond which recovery is impossible is
expected,
but the existence of a lower phase boundary where $q_s$ dominates over
$|\phi_s|$ is novel.
Within the robust phase,
transversal coherent rotations and decoding lead to logical unitary rotations
known to arbitrarily low infidelities.

It is well known that the regimes where quantum error correction is feasible with incoherent noise are described by
ordered phases of classical statistical mechanical models,
whose phase transitions correspond to thresholds%
~\cite{dennisTopologicalQuantumMemory2002,wangConfinementHiggsTransitionDisordered2003,chubbStatisticalMechanicalModels2021}.
To obtain better physical intuition for the emergence of the robust phase, we make use of a recently developed statistical mechanics mapping that can treat mixed coherent-dephasing noise channels  \cite{behrendsStatisticalMechanicalMapping2024,behrendsSurfaceCodePauli2025}. In this case, the statistical mechanics model has complex weights that account for the interference of the error terms from the coherent rotations. 

The key feature of this model is that it consists of two layers of a 2D complex random bond Ising model that are coupled by a four body term arising from the incoherent noise.  At low values of $p$ and $\theta$ the model is in a ferromagnetic phase corresponding to the below threshold regime.  Quenched disorder appears in the model due to the syndrome.   At low-noise rates, but large  values of  $p/\theta$, the two layers of this model are locked together such that the lowest energy excited states consist of a correlated domain wall in the two layers \cite{behrendsSurfaceCodePauli2025}.  However, as $p/\theta \lesssim 1$, the inter-layer coupling becomes weaker and the lowest energy excitation switches to independent domain walls in the two layers.  This energy crossing in the spectrum of the statistical mechanics model is associated with the phase transition to the robust phase [see Fig.~\ref{fig:phase}(a)].  The existence of this robust phase was not noticed in Ref.~\cite{behrendsSurfaceCodePauli2025} because they used a different parameterization for the logical channel that does not readily capture the behavior of the logical relative dephasing parameter.

For sufficiently small target logical angle $\Phi_T$,
the probability of success in the first round is the
probability $p(\vec{0})$ of obtaining the trivial syndrome $s=\vec{0}$ with $\theta$
chosen such that
$\phi_0=\Phi_T$.
Fig.~\ref{fig:phase}(c) shows that the resulting relative dephasing
$q_0/|\phi_0|$ is suppressed exponentially with
distance when exactly targeting the logical angle
$\Phi_T$ at which $p(\vec{0})\approx 50\%$.

\begin{figure*}[th]
    \begin{center}
        \begin{overpic}[scale=1.0]{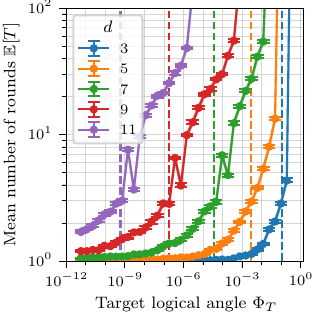}
            \put(0, 100){\textbf{(a)}}
        \end{overpic}
        \hspace{5mm}
        \begin{overpic}[scale=1.0]{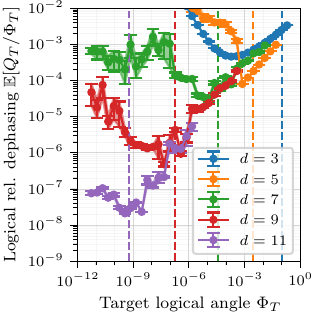}
            \put(0, 100){\textbf{(b)}}
        \end{overpic}
        \hspace{5mm}
        \begin{overpic}[scale=1.0]{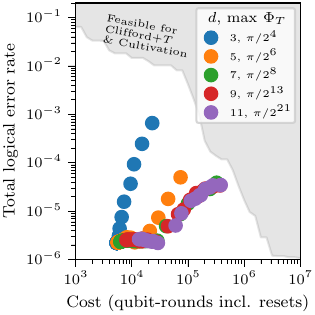}
            \put(0, 100){\textbf{(c)}}
        \end{overpic}
        \vspace{-5mm}
    \end{center}
    \caption{Performance of the adaptive protocol with resets for preparing
    magic states with logical target angle $\Phi_T$ at physical dephasing rate
    $p=0.001$.
    (a) The expected number of rounds $\mathbb{E}[T]$ including resets 
    is independent of the tolerance $\epsilon$, but increases with target angle
    $\Phi_T$.
    (b) There is a target angle where the relative logical dephasing
    $\mathbb{E}[Q_T/|\Phi_T|]$
    is minimal.
    The dashed lines indicate the target $\Phi_T$ at which the protocol will
    succeed in just one round with 50\% probability with the color corresponding
    to the code distance.
    (c) Resource comparison of adaptive protocol against the limits of magic state
    cultivation from \textcite{gidneyMagicStateCultivation2024}
    with Clifford+T synthesis.
    Each point denotes a target angle $\Phi_T$.
    Details are in the Appendix.
    }
    \label{fig:performance}
\end{figure*}

\emph{Adaptive protocol for logical non-Clifford rotations}---%
The robust phase finds application in a fault-tolerant protocol for implementing
non-Clifford logical gates without postselection using only transversal
operations and stabilizer measurements already required for error correction.

Our protocol rotates a surface code state by a target logical $Z$ rotation
angle $\Phi_T$,
while also being fault-tolerant against dephasing errors at a rate $p$.
Transversally rotating all physical qubits by a carefully chosen angle $\theta$,
measuring a syndrome $s$ and applying the correction $C_s$ from the decoder
would result in a logical rotation angle $\phi_s$ and logical dephasing rate
$q_s$,
both of which may be inferred from $s$, $p$, and $\theta$.
As $\phi_s$ may not hit the target $\Phi_T$ on the first try,
we perform more rounds until it does,
tracking the total logical rotation angle $\Phi$
and adaptively choosing $\theta$ for each round
based on a strategy that minimizes $T$, the number of rounds needed.
This is shown in Fig.~\ref{fig:setup}(b).
We can also track the total logical dephasing to assess fault-tolerance,
but keeping $\theta$ in the robust phase ensures that this will be suppressed as
we increase the code distance $d$.

The problem of choosing the best $\theta$ is a classical stochastic control
problem%
~\cite{bertsekasDynamicProgrammingOptimal2012,putermanMarkovDecisionProcesses2014}.
Starting from $\Phi_0=0$,
the system after $t$ rounds of rotation is described by
the total logical rotation angle $\Phi_t$ in the ideal case.
It transitions as $\Phi_{t+1}=\Phi_t + \phi(\theta)$,
where the single-round logical rotation angle $\phi(\theta)$ is a random
variable whose distribution is parameterized by the physical rotation angle
$\theta$,
our tunable control input.
For brevity we have omitted the syndrome $s$ which is the source of randomness.
The cost function to minimize $V(\Phi)$ is the expected number of rounds
$\mathbb{E}[T]$ to
reach the target $\Phi_T$ from the present state $\Phi$ using the optimal
strategy.
The policy prescribing the best input $\theta^*(\Phi)$ for any state
$\Phi$ must satisfy the principle of optimality that the
expected remaining cost to finish should equal the immediate cost $C(\theta)=1$
of doing one round plus the expected remaining cost from the next state.
This is captured by the Bellman equation~\cite{bellmanDynamicProgramming1957}
\begin{align}
    V(\Phi) &= \min_{\theta\in(-\pi/2, \pi/2]}\mathbb{E}\left[ 
        C(\theta) + V(\Phi + \phi(\theta))
    \right]
\end{align}
for $\Phi\ne \Phi_T$ and $V(\Phi_T)=0$ at the target.
The best $\theta$ would minimize the cost for each $\Phi$.
This optimal cost function $V(\Phi)$ may be estimated for a discretized set of
$\Phi$ values by value iteration,
repeatedly evaluating the right hand side using the previous estimate to obtain
a better estimate.
Upon $V(\Phi)$ converging to within numerical tolerance,
the optimal policy $\theta^*(\Phi)$ is obtained by pointwise noting the best
$\theta$ for each $\Phi$ value.

Dephasing introduces an uncertainty of not knowing the operation actually
applied,
resulting in an additional logical dephasing at rate $q(\theta)$ at each step,
which accumulates to a total logical dephasing rate
$Q_{t+1}=(1 - Q_t) q(\theta) + (1 - q(\theta)) Q_t$.

Optimizing for $\Phi_t$ to converge to a target $\Phi_T$ without regards to
$Q_t$ will result in excessively high $Q_t$ values.
If our goal is to prepare a magic state for injection,
we have the option to reset and start afresh.
We can generalize the state space to $(\Phi, Q)$ and optimize the policy
similarly to minimize the total number rounds including resets.
Details of the policy optimization and simulation with dephasing are in the
Appendix.
The expected number of rounds $\mathbb{E}[T]$ needed to achieve the target angle
$\Phi_T$
estimated using this method is plotted in Fig.~\ref{fig:performance}(a).
The mean logical relative dephasing $\mathbb{E}[Q_T/|\phi_T|]$
depends on the physical dephasing rate $p$ and target logical angle $\Phi_T$,
as shown in Fig.~\ref{fig:performance}(b).
As the code distance $d$ increases,
its value is minimized near the target angle $\Phi_T$ at which
one round would succeed with 50\% probability. 

\emph{Discussion}---%
In this Letter,
we have shown the existence of a robust phase of continuous transversal gates
in the surface code.
We exploit this to introduce a low-cost adaptive protocol that efficiently
implements continuously tunable fault-tolerant non-Clifford gates in the surface
code over a range of rotation angles using only transversal operations and
syndrome measurements.
The rotation is exact and postselection was avoided with subsequent rotations
and resets when preparing resource states for injection.
This protocol presents a powerful new addition to the toolkit for fault-tolerant
non-Clifford gates.
Its estimated total logical error rate and cost in terms of expected spacetime
volume including resets for various target angles and distances is plotted in
Fig.~\ref{fig:performance}(c),
where the achievable limits of conventional magic state cultivation and
Clifford+T synthesis are also included for reference.
For example, the $d=7$ protocol becomes competitive when targeting
$\Phi_T \le \pi/2^8$.

Our protocol is complementary to other injection ideas since it natively
generates small angles which other methods struggle to achieve.
However, the range of reachable logical rotation angles decreases exponentially
with increasing code size, which limits the applications to cases where this scaling is desirable.
In particular, the protocol is  more advantageous for algorithms and subroutines requiring
many small rotations,
such as in Trotterized quantum simulation,
where the logical dephasing rate only needs to be below the truncation error%
~\cite{lloydUniversalQuantumSimulators1996}
which depends on the step size proportional to the logical rotation angle%
~\cite{childsTheoryTrotterError2021}.
The quantum Fourier transform~\cite{jozsaQuantumAlgorithmsFourier1998}
used for factoring%
~\cite{shorAlgorithmsQuantumComputation1994,shorPolynomialTimeAlgorithmsPrime1997}
is defined by exponentially small rotations in the input size,
although these may be truncated in practice%
~\cite{coppersmithApproximateFourierTransform2002,barencoApproximateQuantumFourier1996}.
Frameworks such as quantum signal
processing~\cite{lowMethodologyResonantEquiangular2016}
and quantum singular value transform~\cite{gilyenQuantumSingularValue2019}
may also use many small rotations.

A natural extension of this work is to assess performance under
phenomenological and circuit-level noise models with full syndrome
extraction circuits,
for which we expect the results to hold qualitatively by leveraging
existing fault-tolerant protocols of the surface code.
It is expected that the protocol should apply for odd-distance codes satisfying
the logical unitary criterion
in \cite{chengEmergentUnitaryDesigns2024},
which may include more efficient quantum low-density parity check (LDPC) codes%
~\cite{tillichQuantumLDPCCodes2014,panteleevQuantumLDPCCodes2022,leverrierQuantumTannerCodes2022,hastingsFiberBundleCodes2021}.
Further gains may be seen by non-uniform physical qubit rotations,
two-qubit rotations~\cite{ismailTransversalSTARArchitecture2025},
and better decoders.
Experimental demonstration of our protocol will be a future challenge.

\begin{acknowledgements}
\emph{Acknowledgements}---%
We thank Kartiek Agarwal, Benjamin B{\'e}ri,  Matteo Ippoliti, Vedika Khemani, Chris Monroe, and Crystal Noel for helpful discussions. 
  E.H. and M.J.G acknowledge support from Defense Advanced Research Projects Agency (DARPA)
under Agreement No. HR00112490357 and NSF QLCI award no. OMA2120757.  S.G. was supported through the Co-design Center for Quantum Advantage (C2QA) under contract number DE-SC0012704. P.R. acknowledges funding support from NSERC, FRQNT and INTRIQ. EH is supported by the Fulbright Future Scholarship.  This work was performed in
part at the Kavli Institute for Theoretical Physics (KITP), which is supported by grant NSF PHY-2309135.
We are grateful to the authors of \textcite{bravyiCorrectingCoherentErrors2018}
and \textcite{darmawanTensorNetworkSimulationsSurface2017} for sharing
the source code for their algorithms.
Simulation source code will be made available upon publication.
\end{acknowledgements}

\bibliographystyle{apsrev4-2}
\bibliography{papers}

\appendix

\newpage
\emph{Syndrome Sampling}---%
\textcite{bravyiCorrectingCoherentErrors2018} provide an efficient algorithm
to sample the syndrome distribution after single-qubit coherent rotations in an
odd-distance square rotated surface code.
Each surface code physical qubit is treated as the logical qubit of a
\emph{C4 code} encoded over 4 Majorana fermion modes stabilized by a single
generator $S=-c_1 c_2 c_3 c_4$.
The single-qubit Pauli operators $X=i c_1 c_2$ and $Z=i c_1 c_3$ become two-mode
operators,
while surface code stabilizer generators become products of commuting
two-mode terms called \emph{links}.
Initialization, coherent rotations and measurments of these links are fermionic
linear optics on Gaussian states%
~\cite{valiantQuantumComputersThat2001,knillFermionicLinearOptics2001,terhalClassicalSimulationNoninteractingfermion2002},
which are efficiently described by manipulations of a covariance
matrix with only $O(n^2)$ elements $M_{ij}=\Tr[i c_i c_j \rho]$
instead of the $O(2^{2n})$ elements of the density matrix $\rho$.

We extend this to efficiently sample the syndrome distribution with both
coherent rotations and dephasing.
Suppose a configuration of $Z$ errors $Z_e$ described by an $n$-bit string $e$
produces a syndrome $H_X e$ where $H_X$ is the $X$-check matrix.
If the sampled syndrome with coherent rotations only is $s$,
then $H_X e \oplus s$ is the syndrome sampled with rotations and dephasing,
where $\oplus$ denotes bitwise addition modulo 2.

\emph{Logical Channel Calculation}---%
\textcite{darmawanTensorNetworkSimulationsSurface2017} provide a tensor
network algorithm for calculating the logical quantum channel of a rotated
surface code subjected to single-qubit quantum channels on every physical qubit
after measuring a given syndrome.
States may be represented as a projected entangled pair state
(PEPS),
which places a tensor at each qubit with one physical index and virtual indices
contracted against tensors of each neighbouring qubit.
Operators, such as the density matrix, are represented as an projected
entangled pair operator (PEPO) similar to a PEPS but with two physical indices
per tensor.
The projection operator for each stabilizer generator supported on up to 4
qubits is similarly represented with two physical indices per qubit tensor.
To calculate the logical channel parameters via the Choi-Jamiolkowski
isomorphism,
they construct a Bell state
$\ket{\Psi^+}=\ket{\overline{0}}\ket{0} + \ket{\overline{1}}\ket{1}$
between the logical qubit and a perfect ancilla qubit
to compute expectation values of Pauli operators.
The ancilla qubit is represented as an extra physical index of the tensor on a
corner qubit.

The resulting matrix elements of the Choi matrix may then be used to calculate
the channel parameters in Eq. (\ref{eqn:logicalchannel}).
We used exact contraction without truncating the bond dimensions.

\emph{Policy Optimization with Dephasing and Reset}---%
When dephasing is accounted for and reset is allowed,
we start from $(\Phi, Q)=(0, 0)$ to target $(\Phi_T, Q)$ with $Q$ being below
a maximum acceptable logical dephasing $Q_{\textrm{acc}}$ of our choosing so that
$V(\Phi_T, Q)=0$ only for $Q\le Q_{\textrm{acc}}$.
The Bellman equation for non-terminal states becomes
\begin{align}
    V(\Phi, Q) = \min_{\theta\in\mathcal{A}} \mathbb{E} [
    C(\theta) + \gamma V(f( (\Phi, Q), \theta))
    ]
\end{align}
where the action space $\mathcal{A}=(-\pi/2, \pi/2]\cap \set{\textrm{reset}}$
also includes the option to reset at cost $C(\textrm{reset})=1$.
We have also included a discount factor $\gamma$ that may be less than one to
speed up the convergence.
The transition to the next state $f( (\Phi, Q), \theta)$ is defined by
$f( (\Phi, Q), \textrm{reset}) = (0, 0)$ and 
\begin{align}
    f( (\Phi, Q), \theta) &= (\Phi + \phi(\theta), Q + q(\theta) - 2Qq(\theta))
\end{align}
for ordinary physical rotations $\theta\in (-\pi2, \pi/2]$.
We then solve for the optimal policy $\theta^*(\Phi, Q)$
as previously described,
discretizing the state space and using value iteration.

\emph{Simulations with Dephasing}---%
We detail how the results in Fig.~\ref{fig:performance} were obtained.
$N_s=5000$ syndromes were sampled to obtain empirical syndrome distributions
$p(s|\theta)$ for discrete $\theta$ values in the range $0$ to $0.16\pi$.
For intermediate angles, this was linearly interpolated between
those of the nearest available $\theta$ values.
The channel parameters $\log|\phi_s(\theta)|$ and $\log q_s(\theta)$ were
linearly interpolated in $\theta$ with signs.
The policy optimization was done on this parametrized empirical distribution
with the $\Phi$ grid discretized into $N_\Phi=201$ log-spaced bins centered at
the target $\Phi_T$,
and the $Q$ grid discretized into $N_Q=21$ log-spaced bins up to $Q=0.5$.
The action space was discretized into $N_\theta=201$ bins including reset.
The value iteration was done with a discount factor of $\gamma=0.99$,
stopping at a cost function tolerance of $\delta=0.01$.
The time complexity of value iteration is
$O\left( (N_\Phi N_Q)^2 N_\theta \log(1/\delta) / (1 - \gamma) \right)$%
~\cite{bertsekasDynamicProgrammingOptimal2012,putermanMarkovDecisionProcesses2014}.
With resets allowed, the optimization must be done for each target angle
$\Phi_T$.

We simulated the protocol using the optimized policy for 10000 trials by
resampling the interpolated empirical syndrome distribution and
bootstrapping the statistics to get uncertainties,
including the mean number of rounds needed including reset $\mathbb{E}[T]$,
the mean total logical dephasing $\mathbb{E}[Q_T]$
and the mean total relative dephasing $\mathbb{E}[Q_T/\Phi_T]$.

\begin{figure}[h]
    \begin{center}
        \includegraphics[width=\columnwidth]{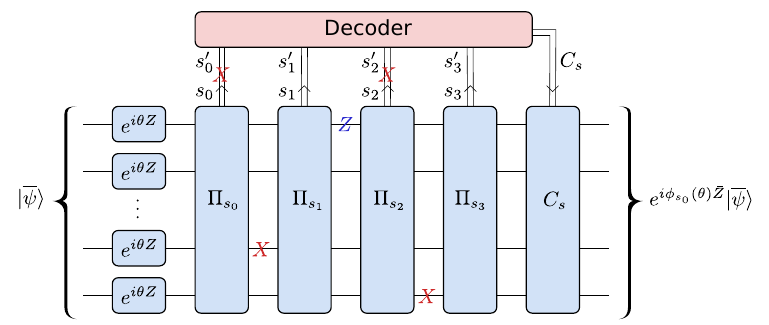}
    \end{center}
    \caption{Repeated rounds of measurement to tolerate measurement faults.}
    \label{fig:ft}
\end{figure}

\emph{Fault-tolerance}---%
To make our protocol tolerate measurement faults,
we may use repeated rounds of syndrome extraction and decoding as shown in
Fig.~\ref{fig:ft}.
The key point is that generic noise after the first syndrome projection does not
affect the logical $Z$ rotation.
This section of the protocol only needs to be below the usual fault-tolerant
threshold for the surface code in order to reliably extract the syndrome result
at the first time slice.
Once this decoding process is completed, one then uses the inferred initial
syndrome to compute the logical angle applied (assuming knowledge of the noise
model).  

\emph{Distribution Properties}---%
We list some useful properties of the syndrome and logical angle distribution.
\begin{enumerate}
    \item The syndrome probability $p(s|\theta)=p(s|-\theta)$
        is an even function of the physical rotation.
        This holds even with a $Z$-error $Z_e$ represented by a bit string $e$.
        $p(s|\theta, e)=p(s|-\theta, e)$.
    \item The logical angle is an odd function of the physical angle.
        $\phi_s(-\theta) = -\phi_s(\theta)$.
        This holds with dephasing
        $\phi_s(-\theta, e) = -\phi_s(\theta, e)$.
    \item Dephasing scrambles the syndrome distribution.
        $p(s|\theta, e)=p(s \oplus H_x e | \theta, \vec{0})$.
    \item
        Dephasing scrambles logical rotation angles and causes logical
        dephasing dependent on the decoder.
        Let $\ell_X\in\mathbb{Z}^n_2$ represent the $\overline{X}$ logical
        operator and let
        $D(s)\in\mathbb{Z}^n_2$
        represent the $Z$-correction $C_s=Z_{D(s)}$ output by the decoder for an
        $X$-syndrome $s$.
        If
        $\ell_X\cdot\left[D(s) + e + D(H_X e)\right] = 0$,
        then
        $\phi_s(\theta, e) = \phi_{s\oplus H_X e}(\theta, 0)$
        otherwise
        $\phi_s(\theta, e) = \phi_{s\oplus H_X e}(\theta, 0) + \pi/2$.
        All bitwise operations are modulo 2 and $\phi_s$ is restricted to 
        $(-\pi/2, \pi/2]$ with periodic boundaries.
\end{enumerate}

\emph{Resource estimation}---%
Let us elaborate the assumptions of Fig.~\ref{fig:performance}(c)
to highlight the regime where our transversal-rotation protocol for preparing
$e^{i\Phi Z}\lvert+\rangle$ may outperform conventional magic-state
cultivation and Clifford+T synthesis,
quantifying cost by spacetime volume including resets and total error rates by
diamond distance.

Conventional gate synthesis of $e^{i\Phi Z}$ requires a $T$-count of
$N_T(\epsilon)\approx 3\log_2(1/\epsilon)$ up to a tolerance $\epsilon$%
~\cite{rossOptimalAncillafreeClifford+T2016}.
Suppose each $\ket{T}$ state may be prepared with an error rate $q(\delta)$
at a spacetime volume cost of $c(\delta)$,
where $\delta$ parameterizes a cost-quality tradeoff such as
in~\cite{gidneyMagicStateCultivation2024}.
Then the total error rate is
$Q_0(\epsilon,\delta)\approx \sqrt{\left(q(\delta) N_T(\epsilon)\right)^2+\epsilon^2}$,
at a total cost of
$C_0(\epsilon,\delta)\approx c(\delta) N_T(\epsilon)$
independent of $\Phi$.
Additionally, magic state injection at each round requires $O(d)$ rounds of
measurements, inccuring a logical error rate of about
$0.1 (100 p)^{(d + 1) / 2}$
per round~\cite{fowlerLowOverheadQuantum2019}.

Our new protocol prepares a logical $e^{i\Phi Z}\lvert+\rangle$ state
with a total logical error rate of
$Q_T(\Phi,d)$
on a distance-$d$
square rotated surface code using $n_q(d)=2d^2-1$ qubits
after $T(\Phi)$ adaptive rounds,
each with $d$ rounds of syndrome extraction,
for a total volume cost of $C_T(\Phi,d)=T(\Phi)n_q(d)d$.
For a fair comparison with state injection,
we assume that each state produced by transversal rotations
is grown to a size $d=11$ code so it may be injected,
which also inccurs an additional logical error rate from $O(d)$ rounds of
measurements at $d=11$.
\end{document}